\documentclass[twocolumn,10pt]{wlscirep}
\title{Ratchet transport powered by chiral active particles}

\author[1,*]{Bao-quan  Ai}
\affil[1]{Guangdong Provincial Key Laboratory of Quantum Engineering and Quantum Materials, School of Physics and Telecommunication
Engineering, South China Normal University, Guangzhou 510006, China}
\affil[*]{corresponding. aibq@scnu.edu.cn}

\begin{abstract}
 \indent  We numerically investigate the ratchet transport of mixtures of active and passive particles in a transversal asymmetric channel.
  A big passive particle is immersed in a 'sea' of active particles.  Due to the chirality of active particles, the longitudinal directed transport
  is induced by the transversal asymmetry. For the active particles, the chirality completely determines the direction of the ratchet transport,
  the counterclockwise and clockwise particles move to the opposite directions and can be separated. However, for the passive particle, the transport behavior becomes complicated, the direction is determined by competitions among the chirality, the self-propulsion speed, and the packing fraction. Interestingly, within certain parameters, the passive particle moves to the left,
  while active particles move to the right. In addition, there exist optimal parameters(the chirality, the height of the barrier, the self-propulsion speed
  and the packing fraction) at which the rectified efficiency takes its maximal value. Our findings could be used for the experimental pursuit of the ratchet transport powered by chiral active particles.
\end{abstract}
\begin{document}
\flushbottom
\maketitle
\thispagestyle{empty}

\section*{Introduction}
\indent Active matter is a rapidly growing branch of nonequilibrium soft matter physics with relevance to
chemistry, biology, and complex systems \cite{rmp1,rpp}. Self-propelled particles are assumed to have an internal propulsion mechanism, which may use energy from an external source and transform it under non-equilibrium conditions into the directed motion. Compared with passive particles, active particles moving in confined structures could exhibit peculiar behaviors\cite{Vicsek1,Nepusz,Cates,Galajda,Leonardo,Wan,Ghosh,Ghosh1,angelani,angelani1,potosky,Potiguar,Guidobaldi,Kaiser1,Koumakis,Costanzo,Ai,Maggi,Yang,McCandlish,Berdakin,Mijalkov,Reichhardt,Buttinoni,Schwarz-Lineka,Schwarz-Lineka1,Fily,MC,Kaiser,Wioland,Rusconi,nc,peruani,Hagen,Ao,Volpe,Zhang,Nguyen,DiLuzio,DiLuzio1,Li,ohta,stark}, resulting for example in collective motion in complex systems\cite{Vicsek1,Nepusz}, spontaneous rectified transport\cite{Cates,Galajda,Leonardo,Kaiser1,Wan,Ghosh,Ghosh1,angelani,angelani1,potosky,Potiguar,Guidobaldi,Koumakis}, separation of active particles based on their swimming properties\cite{Costanzo,Ai,Maggi,Yang,McCandlish,Berdakin,Mijalkov,Reichhardt}, phase separation of self-propelled particles\cite{Buttinoni,Schwarz-Lineka,Schwarz-Lineka1,Fily,MC}, trapping of particles in the microwedge\cite{Kaiser},  spiral vortex formation in the circular confinement\cite{Wioland}, depletion of elongated particles from low-shear regions\cite{Rusconi}, and the other interesting transport phenomena \cite{nc,peruani,Hagen,Ao,Volpe,Zhang,Nguyen,DiLuzio,DiLuzio1,Li,ohta,stark}.

\indent More recently, ratchet effects have been observed in the absence of an external drive for systems of self-propelled particles\cite{Cates}. Experimental studies \cite{Galajda,Leonardo,Kaiser1} show the key role of self-propulsion for rectifying cell motion in an array of asymmetric funnels \cite{Galajda} or for driving a nano-sized ratchet-shaped wheel \cite{Leonardo}. There also has been increasing interest in theoretical work on the rectification of self-propelled particles \cite{Wan,Ghosh,Ghosh1,angelani,angelani1,potosky,Potiguar,Guidobaldi,Koumakis}. The rectification phenomenon of overdamped swimming bacteria was theoretically observed in a system with an array of asymmetric barriers \cite{Wan}.
In a compartmentalized channel, Ghosh and co-workers \cite{Ghosh,Ghosh1} studied the transport of Janus particles  and found that the rectification can be orders of magnitude stronger than that for ordinary thermal potential ratchets. Angelani and co-workers \cite{angelani} studied the run-and tumble particles in periodic potentials and found that the asymmetric potential produces a net drift speed. Potosky and co-workers \cite{potosky}
 found that the spatially modulated self-propelled velocity can induce the directed transport.

\indent In nature and technology, many systems are mixtures of different particle types. Studying and comparing the passive and active motions can provide insight into out-of-equilibrium phenomena\cite{Sergy,Sergy1}. In this paper, we numerically study the directed transport of mixtures of active and passive particles in a transversal asymmetric channel, where the big passive
 is immersed in the 'sea' of active particles. We emphasize on finding how the transversal asymmetry induces the longitudinal directed transport
 and how interactions from the 'sea' of active particles trigger the ratchet transport of the passive particle.

\section*{Model and methods}
\begin{figure}[htbp]
   \vspace{-0.5cm}
   \begin{center}
   \includegraphics[width=0.9\columnwidth]{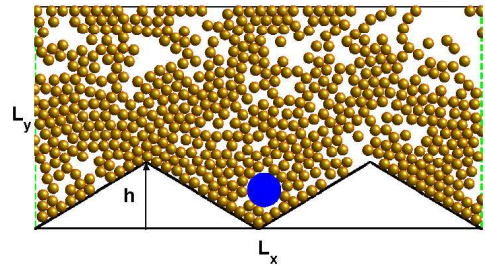}
   \vspace{-1cm}
   \caption{(Color online) Schematic of chirality-powered motor. A M-shaped barrier is regularly arrayed at the bottom of the channel. Periodic boundary
   conditions are imposed in the $x$-direction, and hard wall boundaries in the $y$-direction. The small gold balls denote active particles and the big blue ball denotes the passive particle. }\label{1}                                                                                                                      \end{center}
 \end{figure}

\indent We consider the mixtures of the passive and active particles moving in a two-dimensional channel with hard walls (the height $L_y$) in the $y$-direction and periodic boundary conditions (the period $L_x$) in the $x$-direction (shown in Fig. 1). A M-shape barrier with the height $h$ is regularly arrayed at the bottom of the channel. A big passive particle (blue ball) is immersed in a 'sea' of clockwise (or counterclockwise) particles(gold balls). The dynamics of particle $i$ are described by the position $\mathbf{r}_i\equiv (x_i,y_i)$ of its center and the orientation $\theta_i$ of the polar axis $\hat{\mathbf{n}}_{i}\equiv (\cos\theta_i,\sin\theta_i)$.  The $i$-th particle has radius $a_i$ ($a_i=a_0$ for active particles and $a_i=a_p$ for the passive particle).
The particle $i$ obeys the following overdamped Langevin equations\cite{Mijalkov,Volpe}
\begin{equation}\label{e1}
  \partial_t \mathbf{r}_i =v^0_{i}\hat{\mathbf{n}}_{i}+\mu \sum_{j\neq i}\mathbf{F}_{ij}+\sqrt{2D_{0}}\boldsymbol{\xi}^{T}_{i}(t),
\end{equation}
\begin{equation}\label{e2}
  \partial_t\theta_{i}=\Omega+\sqrt{2D_{\theta}}\xi_{i}(t),
\end{equation}
with $\mu$ the mobility and $v^0_i$ the self-propulsion speed ($v^0_i=v_0$ for active particles and $v^0_i=0$ for the passive particle).
The translational and rotational noise terms, $\boldsymbol{\xi}^{T}_{i}(t)$ and $\xi_{i}(t)$ are Gaussian white noises with zero mean and correlations
$\langle\boldsymbol{\xi}^{T}_{i\alpha}(t)\boldsymbol{\xi}^{T}_{j\beta}(s)\rangle=\delta_{ij}\delta_{\alpha\beta}\delta(t-s)$ ($\alpha$, $\beta$
labels denote Cartesian coordinates)and  $\langle \xi_{i}(t)\xi_{j}(s)\rangle = \delta_{ij}\delta(t-s)$.
$\langle...\rangle$ denotes an ensemble average over the distribution of noise and $\delta$ the Dirac delta function. $D_{0}$ and $D_{\theta}$ denote the
translational and rotational diffusion coefficients, respectively. $\Omega$ is the angular velocity and its sign
determines the chirality of active particles.
We define particles as  the clockwise particles for negative $\Omega$ and the counterclockwise particles for positive $\Omega$.

 \indent The force $\mathbf{F}_{ij}$ between particles $i$ and $j$ is assumed to be of the linear spring form with the stiffness constant $k$:
$\mathbf{F}_{ij}=F_{ij}\hat{\mathbf{r}}_{ij}$, with $\hat{\mathbf{r}}_{ij}= (\mathbf{r}_i-\mathbf{r}_j)/r_{ij}$, $r_{ij}=|\mathbf{r}_i-\mathbf{r}_j|$, and $F_{ij}=k(a_i+a_j-r_{ij})$ if $r_{ij}<a_i+a_j$ and $F_{ij}=0$ otherwise. The interactions between particles are radially symmetric and do not directly coupled to angular dynamics.
We define the ratio between the area occupied by particles and the total available area as the packing fraction $\phi=\sum_{i=1}^{N}\pi a_i^2/(L_x L_y-\frac{1}{2}L_x h)$, where $N$ is the total number of particles.  For the high packing fraction $\phi$ in excess of 1, particles of either species overlap on average.

\indent Eqs.(\ref{e1},\ref{e2}) can be rewritten in the dimensionless forms by introducing characteristic length scale
and time scale: $\hat{\mathbf{r}}=\frac{\mathbf{r}}{a_0}$, $\hat{t}=\mu k t$,
\begin{equation}\label{e3}
  \partial_{\hat{t}} \hat{\mathbf{r}}_i =\hat{v}^{0}_i\hat{\mathbf{n}}_{i}+\sum_{j\neq i}\hat{\mathbf{F}}_{ij}+\sqrt{2\hat{D}_{0}}\boldsymbol{\hat{\xi}}^{T}_{i}(\hat{t}),
\end{equation}
\begin{equation}\label{e4}
  \partial_{\hat{t}}\theta_{i}=\hat{\Omega}+\sqrt{2\hat{D}_{\theta}}\hat{\xi}_{i}(\hat{t}),
\end{equation}
and the other parameters can be rewritten as $\hat{v}_0=\frac{v_0}{\mu k a_0}$, $\hat{\Omega}=\frac{\Omega}{\mu k}$, $\hat{D}_{\theta}=\frac{D_{\theta}}{\mu k}$, $\hat{D}_{0}=\frac{D_0}{\mu k a^{2}_{0}}$, $\hat{a}_p=\frac{a_p}{a_0}$, $\hat{L}_{x,y}=\frac{L_{x,y}}{a_0}$, and $\hat{h}=\frac{h}{a_0}$. From now on, we will use only dimensionless variables and shall omit the hat for all quantities occurring in the above equations.

\indent The behaviors of quantities of interest can be corroborated by Brownian dynamic simulations performed by the integration of Langevin equations(\ref{e3},\ref{e4}). We only consider the $x$-direction average velocity because particles are confined in the $y$-direction.
The average velocity along the $x$-direction in the asymptotic long-time regime can be obtained from the formula
$\langle V_x\rangle=\frac{1}{N}\sum_{i=1}^{N}\lim_{t\rightarrow\infty}\frac{[x_i(t)-x_i(0)]}{t}$. We define the scaled average velocity $V_s=\langle V_x\rangle/v_0$ for convenience.
\section*{Results and Discussion}
\indent For numerical simulations, the total integration time was more than $10^{7}$ and the transient effects were estimated
and subtracted. The integration step time $\Delta t$ was chosen to be smaller than $10^{-4}$.  With these parameters, the simulation results are robust.
Unless otherwise noted, our simulations are under the parameter sets: $D_0=10^{-4}$, $D_\theta=5\times10^{-4}$, $L_x=40.0$, $L_y=20.0$, and $a_p=3.0$.

\indent  As we know, in nonlinear systems, the ratchet setup demands two key ingredients\cite{ratchet} which are (a)Fluctuating
input zero-mean force: it should break the thermodynamical equilibrium, which forbids appearance of the directed transport due to the Second Law of Thermodynamics. (b)Asymmetry (temporal and/or spatial): it can violate the left-right symmetry of the response. For our system,
the term $v_0 \cos\theta$ in Eq.(\ref{e1}) can be seen as the fluctuating input zero-mean force and the asymmetry comes from the upper-lower asymmetry of the channel. Now we will discuss how the chirality of active particles breaks the thermodynamical equilibrium and induces the ratchet transport.

\begin{figure}[htbp]
\vspace{0cm}
\begin{center}
\includegraphics[width=0.9\columnwidth]{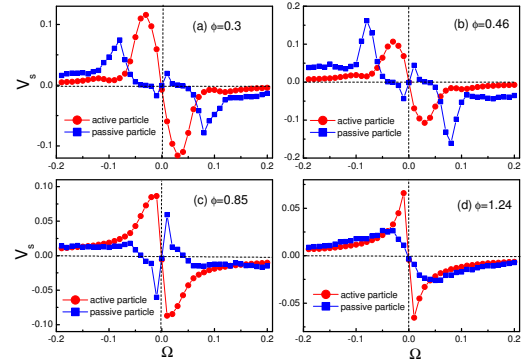}
\caption{(Color online) Average velocity $V_s$ as a function of the angular velocity $\Omega$. (a)$\phi=0.3$. (b)$\phi=0.46$. (c)$\phi=0.85$. (d)$\phi=1.24$.
 The other parameters are $v_0=0.5$ and $h=10.0$.}\label{1}
\end{center}
\end{figure}
\indent Figure 2 shows the average velocity $V_s$ as a function of the angular velocity $\Omega$ for different values of $\phi$. For active particles (denoted by the red lines), $V_s$ is negative for $\Omega>0$, zero at $\Omega=0$, and positive for $\Omega<0$. The movement direction of active particles is completely determined by the sign of $\Omega$. In other words, active particles with different chiralities move to different directions and can be separated. In addition, when $|\Omega|\rightarrow \infty $, the self-propelled angle changes very fast, particles will experience a zero averaged force,
so $V_s$ tends to zero.  Therefore, there exists an optimal value of $|\Omega|$ at which $|V_s|$ takes its maximal value.

\indent Now we explain the rectified mechanism of chiral active particles in the upper-lower asymmetric channel(see Fig. 3). In a wide channel,
if the channel cell is large enough and no external perturbations, chiral active particles will perform the circular motion and repeat it and the radius of the circular trajectory is about $v_0/|\Omega|$.  However, in our system, the radius of the circular trajectory is much larger than the channel cell
and chiral particles can not perform circular motion. Due to the confinement of the channel, particles slide along the walls. For the clockwise particles ($\Omega<0$) shown in Fig. 3, due to the upper-lower asymmetry of the channel, the motion time along the lower wall is significantly larger than along the upper wall, therefore, the clockwise particles on average move to the right. In a similar way, the counterclockwise particles ($\Omega>0$) will on average move to the left.

 \indent For the single passive particle, the ratchet effect disappears ($V_s=0$) due to the absence of the fluctuating input zero-mean force. However, when the passive particle was immersed in the 'sea' of active particles, the interactions from active particles can break the thermodynamical equilibrium and make the passive particle move directionally.  At the high packing fraction (e. g. $\phi=1.24$ shown in Fig. 2(d)), the passive and active particles have the similar transport behaviors. This is because the passive particle is completely controlled by chiral active particles at the high packing fraction.
 However, when $\phi<1.0$, the ratchet behaviors of the passive particle (blue lines) are different from those of active particles (red lines).
 Interestingly, for very small values of $|\Omega|$ (e. g.  $|\Omega|=0.01$), the passive and active particles move to the opposite directions. Since the transport behavior powered by the counterclockwise particles is completely opposite to that powered by the clockwise particles, we only consider the case of the clockwise particles ($\Omega<0$) in the following discussion.
\begin{figure}[htbp]
\vspace{2cm}
\begin{center}
\includegraphics[width=0.5\columnwidth]{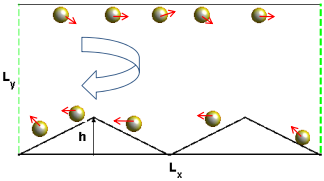}
\vspace{0cm}
\caption{(Color online)Sketch of the rectified mechanism of clockwise particles in the upper-lower asymmetric channel.}\label{1}
\end{center}
\end{figure}

\begin{figure}[htbp]
\begin{center}\includegraphics[width=0.9\columnwidth]{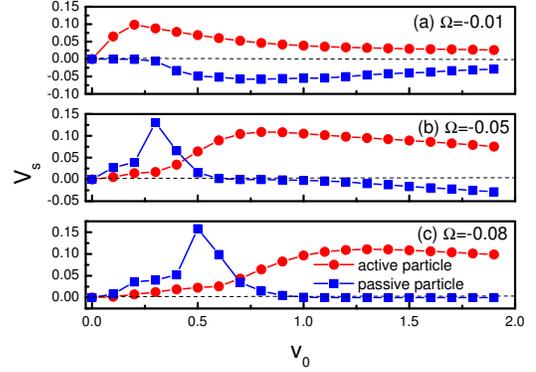}
\caption{(Color online) Average velocity $V_s$ as a function of the self-propulsion speed $v_0$ for both active and passive particles. (a)$\Omega=-0.01$. (b)$\Omega=-0.05$. (c)$\Omega=-0.08$. The other parameters are $\phi=0.46$ and $h=10.0$. }\label{1}
\end{center}
\end{figure}

\indent In Fig. 4 we present the average velocity $V_s$ as a function of the self-propulsion speed $v_0$ for both active and passive particles.
For the case of active  particles (see the red lines in Fig. 4), all curves are observed to be bell shaped, and there exists an optimal value
of $v_0$ at which $V_s$ takes its maximal value. When $v_0$ tends to zero, the fluctuating input disappears and the ratchet effect disappears,
thus $V_s$ is nearly equal to zero. For very large values of $v_0$, the chirality of the particle can be negligible, the left-right symmetry can not be broken and the directed transport gradually disappears. Therefore, the optimal self-propulsion speed can facilitate the rectification of active particles.  As $|\Omega|$ increases, the position of the peak shifts to the large values of $v_0$.

\indent For the case of the passive particle (see the blue lines in Fig. 4), the transport behavior becomes complicated. Similar to Figs. 2(a)-2(c),
 current reversals occur for the case of very small value of $|\Omega|$ (e. g. $\Omega=-0.01$). The average velocity is always positive for the case of $\Omega=-0.08$. Interestingly, for the case of $\Omega=-0.05$, the passive particle moves to the right for  $v_0<1.0$ and the left for $v_0>1.0$. Therefore, the self-propulsion speed can also determine the movement direction of the passive particle.

\begin{figure}[htbp]
\vspace{-1cm}
\begin{center}\includegraphics[width=0.9\columnwidth]{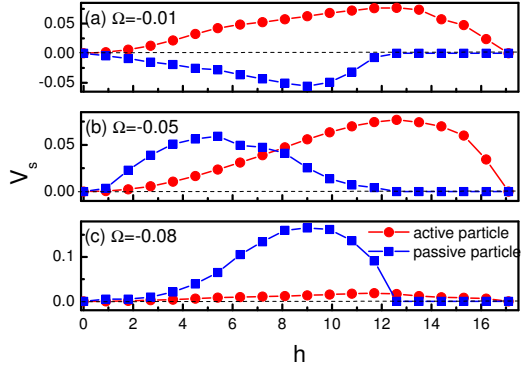}
\caption{(Color online)Average velocity $V_s$ as a function of the height $h$ of the M-shaped barrier for both active and passive particles. (a)$\Omega=-0.01$. (b)$\Omega=-0.05$. (c)$\Omega=-0.08$.  The other parameters are $\phi=0.46$ and $v_0=0.5$.}\label{1}
\end{center}
\end{figure}

\indent The dependence of the average velocity $V_s$ on the height $h$ of the M-shaped barrier is shown Fig. 5. For both passive and active particles,
 all curves are observed to be bell shaped, and there exists an optimal value of $h$ at which $|V_s|$ takes its maximal value.
 When $h\rightarrow 0$, the asymmetry will disappear and no directed transport occurs. For very large values of $h$, the channel is blocked,
 particles cannot pass through the M-shape barrier, thus $v_s$ tends to zero. Therefore, the optimal height can facilitate the ratchet transport.

\begin{figure}[htbp]
\vspace{-1cm}
\begin{center}\includegraphics[width=0.9\columnwidth]{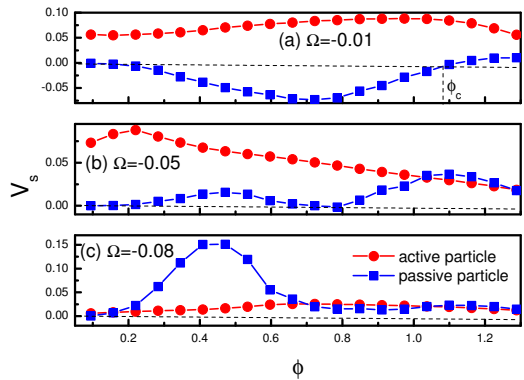}
\caption{(Color online) Average velocity $V_s$ a function of the packing fraction $\phi$ for both active and passive particles. (a)$\Omega=-0.01$. (b)$\Omega=-0.05$. (c)$\Omega=-0.08$. The other parameters are $v_0=0.5$ and $h=10.0$.}\label{1}
\end{center}
\end{figure}
\indent  Figure 6 shows the average velocity $V_s$ as a function of the packing fraction $\phi$ for both active and passive particles. For the case of active particles (see the red lines in Fig. 6), the average velocity $V_s$ is a peaked function of $\phi$. We will then explain this behavior. The interactions between particles can cause two results: (A) reducing the self-propelled driving, which blocks the ratchet transport and (B)activating motion in an analogy with the thermal noise activated motion for a single stochastically driven ratchet, which facilitates the ratchet transport. When the packing fraction increases from zero, the factor B first dominates the transport, so the average velocity increases with the packing fraction. However, when the packing faction become large, the factor A dominates the transport, thus the average velocity decreases with increasing $\phi$.  Therefore, there exists an optimal value of $\phi$ at which the average velocity is maximal.

\indent For the passive particle ((see the blue lines in Fig. 6), the transport behavior ($V_s$ vs $\phi$) becomes more complicated. When $|\Omega|>0.05$ (e. g. $\Omega=-0.08$), $V_s$ is positive and there exist two peaks in the curve. For very small values of $|\Omega|$ (e. g. $\Omega=-0.01$),
$V_s$ is negative for $\phi<\phi_c$, zero at $\phi=\phi_c$, and positive for $\phi>\phi_c$. Therefore, we can also have current reversals by changing the packing fraction.

\section*{Concluding remarks}

\indent In conclusion, we numerically studied the transport of mixtures of active and passive particles moving in a periodic channel
  with a M-shaped barrier. A big passive particle was immersed in the 'sea' of active particles.  The longitudinal ratchet transport of particles can be induced by the transversal asymmetry. The interactions from chiral active particles can make the passive particle move directionally. For active particles, the direction of the ratchet transport is completely determined by the chirality of active particles, the average velocity is positive for $\Omega<0$, zero at $\Omega=0$ and negative for $\Omega>0$.
 In other words, the counterclockwise and clockwise particles move to the opposite directions and can be separated.
 However, the transport behavior of the passive particle becomes complicated, the direction of the ratchet transport is determined by competitions among
 the chirality, the self-propulsion speed, and the packing fraction. Remarkably, within certain parameters, the passive and active particles
 move to the opposite directions, for example, the big passive particle moves to the left, while active particles move to the right
 when $\Omega=-0.01$ and $\phi<1.0$. We also found that there exist optimal parameters (the chirality, the height of the barrier, the self-propulsion speed and the packing fraction) at which the average velocity takes its maximal value.
 \indent  Our results should be of considerable practical and theoretical interest, because they provide new insights into
 active matter and non-equilibrium systems. Applications of these results can be envisioned for ion mixtures traveling through cell
 membranes or moving through artificial nanopores, for the controlling transport in colloidal suspensions, and for the particle separation.

\section*{Acknowledgements}
\indent This work was supported in part by the National Natural Science Foundation of China (Grant Nos. 11575064 and 11175067), the PCSIRT (Grant No. IRT1243),  the Natural Science Foundation of Guangdong Province (Grant No. 2014A030313426), and Program for Excellent Talents at the University of Guangdong Province.

\section*{Author contributions statement}
B. Q. Ai. carried out the numerical modeling and wrote the paper.
\section*{Additional information}
\textbf{Competing financial interests:} The authors declare no competing financial interests.
\end{document}